\documentstyle[12pt,epsf]{article}
\begin{document}
     
%
     
\newcommand\ie {{\it i.e.}}
\newcommand\eg {{\it e.g.}}
\newcommand\etc{{\it etc.}}
\newcommand\cf {{\it cf.}}
\newcommand\etal {{\it et al.}}
\newcommand{\be}{\begin{eqnarray}}
\newcommand{\ee}{\end{eqnarray}}
\newcommand{\jp}{$ J/ \psi $}
\newcommand{\pp}{$ \psi^{ \prime} $}
\newcommand{\ppp}{$ \psi^{ \prime \prime } $}
\newcommand{\dd}[2]{$ #1 \overline #2 $}
\newcommand\noi {\noindent}

\begin{flushright}
LBNL-47716
\end{flushright}
\vspace{1cm}
     
\begin{center}
     
{\Large {Are the $J/\psi$ and $\chi_c$ $A$ Dependencies the Same?}}
\vspace{2mm}

R. Vogt\footnote{This work was supported in part by
the Director, Office of Energy Research, Division of Nuclear Physics
of the Office of High Energy and Nuclear Physics of the U. S.
Department of Energy under Contract Number DE-AC03-76SF00098.} \\
\vspace{2mm}
{\it Niels Bohr Institute, Blegdamsvej 17, DK-2100 Copenhagen} \\
{\it Nuclear Science Division, Lawrence Berkeley National Laboratory, 
Berkeley, CA 94720, USA} \\
{\it Physics Department, University of California at Davis, Davis, CA 95616, 
USA}\\
\vspace{2mm}

     
\vspace{1cm}
{\bf Abstract}\\
\begin{quote} \begin{small}
It has been empirically observed that the dependence of $J/\psi$ and $\psi'$ 
production on nuclear mass number $A$ is very similar.  This has been 
postulated to be due to the predominance of color octet
pre-resonant states in charmonium production and absorption.  
Two new experiments, NA60 at CERN and HERA-B at DESY, will
measure the $\chi_c$ $A$ dependence for the first time.  These measurements 
should shed new light on the charmonium production and absorption mechanisms.
\\[2ex] 
\end{small} \end{quote} 
PACS: 24.85.+p, 25.40.Ep
\end{center}
\newpage

\section{Introduction}

The description of particle production in proton-nucleus interactions as a
simple scaling of the proton-proton production
cross section with nuclear mass number $A$,
\be
\sigma_{pA} = \sigma_{pN} A^\alpha \, \, , \label{alfint}
\ee 
has been used to describe many processes.  When the production of the desired
final state particle is calculable in perturbative QCD, the
factorization theorem \cite{CSS} suggests that the exponent $\alpha$ should be
unity.  Drell-Yan production, integrated over all kinematic variables,
agrees with $\alpha=1$ to rather high precision
\cite{E772DY} although some deviation from unity appears at large values of
Feynman $x$, $x_F = p_\parallel/p_{\rm max}$.  
A less than linear $A$ dependence has been observed for $J/\psi$, $\psi'$, 
$\Upsilon$, and $\Upsilon' + \Upsilon''$ production
with $0.9< \alpha < 1$ near $x_F 
\approx 0$. 

By now, the $A$ dependence of $J/\psi$ production at $x_F >0$
is known to rather high precision at several different energies
\cite{NA3,E537,E772,E789,na50,e866}.  While the 
$\psi'$ $A$ dependence is not as accurately known, its statistics 
were sufficient for the E866 collaboration to determine that the $\psi'$ 
$\alpha$ is smaller than the $J/\psi$ $\alpha$ for
$x_F < 0.2$ \cite{e866}.  The known $A$ dependence of $J/\psi$ production
has been used to determine the strength of the ``anomalous'' $J/\psi$ 
suppression in Pb+Pb interactions at the CERN SPS \cite{na50qgp}.  However,
an important assumption in this interpretation is that all
charmonium states interact with the nucleus while in ``pre-resonant'' $|(c 
\overline c)_8 g \rangle$ states \cite{KSghfit}.  
Since a significant fraction,
$\sim 40$\%, of the observed $J/\psi$'s come from $\chi_c$ decays 
\cite{Kirk,Kourk,Clark,Hahn,Bauer,Koreshev,Lemoigne,Ant1,Ant2,Antonxf,Ronceux},
a measurement of the $\chi_c$ $A$ dependence is crucial for the understanding
of $J/\psi$ suppression in nucleus-nucleus collisions because in a quark-gluon
plasma, $J/\psi$ suppression is expected to occur in steps, the first of
which is the dissociation of the $\chi_c$ \cite{KMS}.  So far, only one
experiment \cite{Antonxf} has presented differential distributions of $\chi_c$
production, albeit with such a small sample that it is not possible to tell
if the shapes of the $\chi_c$ and $J/\psi$ $x_F$ distributions are the same
or not.  No measurement of the $\chi_c$ $A$ dependence has yet been made.

Fortunately, this situation seems about to change.  The $\chi_c$
$A$ dependence will be measured for the first time in two fixed-target
experiments at
different energies.  The NA60 collaboration, a follow-up to the NA50 
collaboration at CERN, has been approved for $pA$ measurements at 450 GeV and
is planning to also take data in nucleus-nucleus interactions at 158 GeV 
\cite{NA601,NA602}.  Their muon spectrometer will sit at $0< y_{\rm cm} < 1$ at
both energies, giving forward
$x_F$ coverage only.  The HERA-B collaboration at DESY has placed target wires
around the halo of the proton beam at HERA.  In their first run, they have
demonstrated that they can detect the
$\chi_c$ \cite{HERAB1,HERAB2}.  In their next run,
they will measure the $J/\psi$, $\psi'$, and $\chi_c$ $A$ dependence over 
$-0.5<x_F<0.3$.  This will thus be 
the first charmonium experiment with coverage 
significantly below $x_F \sim -0.1$.

If the $A$ dependence of $\chi_c$ production is the same as that of 
the $J/\psi$,
then the picture of a pre-resonant color octet state passing through the target
\cite{KSghfit} will be validated.  
Then charmonium production and absorption at fixed-target energies can
be essentially described within the color evaporation model \cite{HPC}.
However, the nonrelativistic QCD model \cite{bbl,benrot}
predicts that $\chi_c$ production should
be predominantly color singlet while direct $J/\psi$ and $\psi'$ production
is via color octet states.  If this picture is correct, 
the $A$ dependence of $\chi_c$ production could
be quantitatively different than that of the $J/\psi$ and $\psi'$. 

In this paper we focus only on charmonium production and its subsequent 
absorption by 
nucleons, the ``normal absorption'' identified by NA50 \cite{na50qgp}.  While
this is insufficient to describe $\alpha(x_F)$ over the full range of $x_F$,
the NA60 and HERA-B
measurements will be in a region where the $x_F$ dependence of
$\alpha$ has so far either not been determined or has
not been strong \cite{e866}.  At larger negative $x_F$, the
$A$ dependence may be different than expected from pre-resonant absorption
\cite{KSghfit}.
Other nuclear effects such as shadowing, 
energy loss, and intrinsic heavy 
quarks depend only on either the projectile or target momentum fractions and
not on the identity of the final charmonium state \cite{RVPRC} and thus
should affect $J/\psi$ and $\chi_c$ production identically.
We first discuss charmonium production by color evaporation and nonrelativistic
QCD and then describe how nuclear absorption of color octet and color singlet 
states might be disentangled.
 
\section{Charmonium Production: Color Evaporation vs. NRQCD}

Two models have been used to describe quarkonium hadroproduction: the
color evaporation model (CEM) and the nonrelativistic QCD model (NRQCD).
Since both have been described in detail elsewhere, we only discuss the 
specifics that are germane to our calculation.

In the CEM, charmonium production is a subset of the $c \overline c$ pairs
produced below the $D \overline D$ threshold.  The
hadronization of charmonium state $C$ from these sub-threshold 
$c \overline c$ pairs is
accomplished through the emission of one or more soft gluons.
It is assumed that the
hadronization does not affect the kinematics of the parent
$c \overline c$ pair so that only a 
single universal factor, $F_C$, is necessary for each state.
The factor $F_C$ depends on the charm quark mass, $m_c$, 
the scale $\mu$ of the strong coupling constant $\alpha_s$, and the 
parton densities.  We use the 
MRST LO parton distributions \cite{mrsg1,mrsg2} for CEM production.  
The factor $F_C$ must be 
constant for the model to have any predictive power.
The differential and
integrated quarkonium production rates should thus be proportional to each 
other and independent of
projectile, target, and energy.  The relative charmonium rates so far seem to
bear this out since $\sum_J \chi_{cJ}/ (J/\psi) \approx 0.4$ 
and $\psi^\prime/(J/\psi) \approx
0.14$ \cite{Ant1,Ant2,Ronceux,Teva,CarlosEPS} over a wide range of targets
and energies, see also Ref.~\cite{HPC}.

The LO cross section of state $C$, 
$\tilde{\sigma}_C$, from projectile $p$ and target $A$
is \be \frac{d \tilde{\sigma}_C}{d x_F} & = &
2F_C^{\rm NLO}
K \int_{2m_c}^{2m_D} m \, dm \, \int_0^1 dx_1 dx_2 \, \delta(x_1x_2 s - m^2) \,
\delta ( x_F - x_1 + x_2 ) \nonumber \\
&  &\mbox{} \times \bigg\{ f_g^p(x_1,m^2)f_g^A(x_2,m^2)\sigma_{gg}(m^2)
  \nonumber \\    & & \mbox{} 
+ \sum_{q=u,d,s} [f_q^p(x_1,m^2) f_{\overline q}^A(x_2,m^2) + f_{\overline 
q}^p(x_1,m^2) f_q^A(x_2,m^2)] \sigma_{q \overline q}(m^2) \bigg\} 
 \, \, . \label{cevap} \ee  
The partonic cross sections $\sigma_{gg}$  and $\sigma_{q \overline q}$ can
be found in Ref.~\cite{Bkp21,Bkp22}.  
Production by quark-gluon scattering enters only at NLO.

A $K$ factor was included in Eq.~(\ref{cevap}) since our calculation is at
leading order and $F_C$ was determined at next-to-leading order, as indicated.
At NLO, the charmonium cross section
was calculated using the $Q \overline Q$ production code of 
Ref.\ \cite{MNR} with a cut on the pair mass as in Eq.~(\ref{cevap}) 
\cite{HPC}.  The $p_T$ dependence and the normalization of the charmonium cross
section from the Tevatron 
collider agrees with these
calculations \cite{SchulV}.  Since, at fixed energy, the $K$ factor for $c
\overline c$ production is 
independent of the kinematic variables \cite{RVZPC}, 
our calculation is at leading order.
Therefore, we multiply the LO cross section by $K$ to obtain the magnitude of
the NLO cross section and then also 
by $F_C^{\rm NLO}$ to fix the hadronization of 
the subthreshold $c \overline c$ pairs to charmonium.  Note however that since
we study ratios of cross sections, only the relative normalization is important
and because no nuclear effects on the parton densities are included, the CEM
production information generally cancels.

Since the CEM depends on the universality of charmonium hadronization through
soft gluon emission, a check of this assumption for $\chi_c$ production, 
particularly as a function of $x_F$, is critical.  The $\chi_c$ has previously
been crucial for furthering the understanding of charmonium production.  The
color singlet model (CSM) \cite{baru1,baru2} described high $p_T$ charmonium
production as direct color singlet production with the appropriate quantum
numbers.  In the CSM, direct $J/\psi$ and $\psi'$ production required the
emission of a hard gluon and should thus be rare on a perturbative timescale.
However, $\chi_c$'s could be directly produced as color singlets and thus high
$p_T$ $J/\psi$ production should be dominated by $\chi_c$ decays.  The
measurement of $\chi_c$ relative to direct $J/\psi$ production at the Tevatron
collider \cite{ppbar} showed that the CSM was incomplete.

The non-relativistic QCD 
approach to quarkonium production was formulated \cite{bbl} was formulated as
a way to go beyond the CSM.
NRQCD describes quarkonium production as an expansion in powers of $v$, 
the relative $Q$-$\overline Q$ velocity.  Thus 
the angular momentum or color of the quarkonium state is not restricted 
to only the leading color singlet state but includes color
octet production as well. 

The $x_F$ distribution of charmonium state $C$ in NRQCD is
\be  \frac{d \sigma_C}{dx_F} & = & \sum_{i,j}  \sum_n
\int_0^1 dx_1 dx_2 \delta ( x_F - x_1 + x_2 ) f_i^p(x_1,\mu^2)f_j^A(x_2,\mu^2)
C^{ij}_{c \overline c \, [n]} 
\langle {\cal O}_n^C \rangle \, \, , \label{signrqcd} \ee 
where the partonic cross section
is the product of perturbative expansion coefficients, 
$C^{ij}_{c \overline c \, [n]}$,
and nonperturbative parameters describing the hadronization,
$\langle {\cal O}_n^C \rangle$.  We
use the parameters determined by Beneke and Rothstein for fixed-target
hadroproduction using the CTEQ 3L parton densities \cite{cteq3}
with $m_c = 1.5$ GeV and $\mu = 2m_c$ \cite{benrot}.  Since the parameters
$\langle {\cal O}_n^C \rangle$ are fit to the LO calculation with a LO set of
parton densities, no further $K$ factor is required.

Direct $J/\psi$ production has only contributions
from $gg$ fusion and $q \overline q$ annihilation \cite{benrot}, as in the CEM.
The $q \overline q$ contribution is all octet while the $gg$ component is a
combination of octet and singlet production.  The $gg$ partonic
cross sections for $J/\psi$ and $\psi'$ production are
\be \widehat{\sigma}(gg \rightarrow \psi) = C^{gg}_{c \overline c \, [n]} 
\langle {\cal O}_n^\psi \rangle = B_8(x_1,x_2,s,m_c^2)
\Delta_8(\psi) + B_1(x_1,x_2,s,m_c^2) \langle {\cal O}_1^\psi ( ^3S_1 )
\rangle \label{psigg}
\ee
where only the octet, $\Delta_8^\psi = \langle {\cal O}_8^\psi (
^1S_0) \rangle + (7/m_c^2)\langle {\cal O}_8^\psi ( ^3P_0) \rangle$, and 
singlet, $\langle {\cal O}_1^\psi ( ^3S_1) \rangle$, matrix elements 
differ between $J/\psi$ and $\psi'$ production.  The functions
$B_1$ and $B_8$ are proportional to $\alpha_s^2$ and $\alpha_s^3$ respectively.
The octet parameters
are quite different for the two states: $\Delta_8(J/\psi) 
\approx 5.8 \Delta_8(\psi')$.  The smaller $\Delta_8(\psi')$ could
be due to the larger mass and thus the increased
``hardness'' of the emitted gluon for the $\psi'$.

On the other hand, a color singlet $\chi_c$ can be formed from two gluons 
\cite{baru1,baru2} 
so that $\chi_c$ production is predominantly color singlet.  In 
addition, the $\chi_{c1}$ has a singlet contribution from $gq$ scattering at
${\cal O}(\alpha_s^3)$ \cite{benrot}.  Only the $q \overline q$ channel 
contributes to color octet $\chi_c$ production.  Thus 
the largest singlet contribution to total $J/\psi$ production is from
$\chi_{cJ}$ decays.  Of these $\chi_{cJ}$ decays, 
the most important is the $\chi_{c1}$ which 
has a 27\% branching ratio to $J/\psi$.  The $\chi_{c2}$ also has a relatively
large branching ratio to $J/\psi$, 14\%.
Although the $\chi_{c0}$ production cross section is as large as those of the 
other $\chi_c$ states, its small branching ratio, $<1$\%, 
results in a negligible
$\chi_{c0}$ contribution to $J/\psi$ production.  The $\chi_{c0}$ is
essentially invisible
in hadroproduction experiments which reconstruct $\chi_{cJ}$'s from their
radiative decays to $J/\psi$.  

\begin{figure}[htbp]
\setlength{\epsfxsize=\textwidth}
\setlength{\epsfysize=0.55\textheight}
\centerline{\epsffile{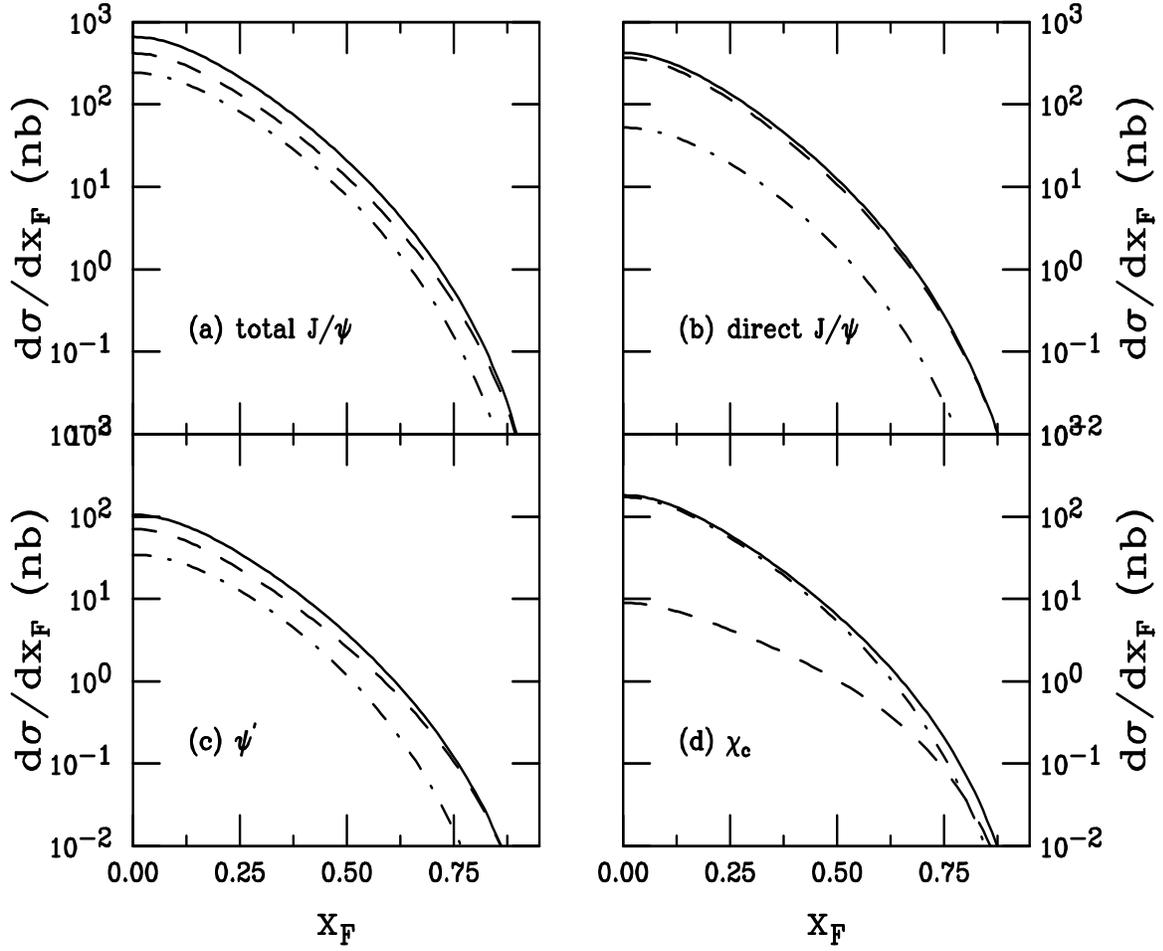}}
\caption[]{Charmonium $x_F$ distributions at 450 GeV. The total $J/\psi$ (a),
direct $J/\psi$ (b), $\psi'$ (c) and summed $\chi_{cJ}$ contributions to
the $J/\psi$ (d) cross sections
are shown.  The octet (dashed) and singlet (dot-dashed) contributions to the
total (solid) are shown separately.} 
\label{nrqcd}
\end{figure}

The total $J/\psi$ $x_F$ distribution then includes radiative decays of the
three $\chi_{cJ}$ states and hadronic decays of the $\psi'$,
\be \frac{d \sigma_{J/\psi}}{dx_F} =  \frac{d \sigma_{J/\psi}^{\rm dir}}{dx_F}
+ \sum_{J=0}^2 B(\chi_{cJ} \rightarrow J/\psi X) \frac{d 
\sigma_{\chi_{cJ}}}{dx_F}
+ B(\psi' \rightarrow J/\psi X) \frac{d\sigma_{\psi'}}{dx_F} \label{dirpsi}
\, \, . \ee  In Fig.~\ref{nrqcd} we show an example of the relative singlet and
octet contributions to total $J/\psi$, direct $J/\psi$, $\psi'$ and the sum
of the three $\chi_c$ contributions to $J/\psi$ production at 450 GeV, the SPS
proton beam energy.  Only the forward $x_F$ distributions are shown since the
distributions are symmetric around $x_F= 0$.  No nuclear effects on the
parton distribution functions are included.

The percentage octet production of each charmonium state 
is given in Table~\ref{octfrac}.
\begin{table}
\begin{center}
\begin{tabular}{|c|c|c|c|c|} \hline
$\sqrt{s}$ (GeV) & Total $J/\psi$ (\%) & Direct $J/\psi$ (\%) & 
$\,\,\,\psi'\,\,\,$ (\%)
& $\sum_J \chi_{cJ} \rightarrow J/\psi$ (\%)
\\ \hline
17.3 & 66.6 & 90.7 & 75.2 &  8.9 \\ \hline
29.1 & 62.6 & 86.7 & 66.2 &  6.3 \\ \hline
41.6 & 60.4 & 84.7 & 61.9 &  5.0 \\ \hline 
\end{tabular}
\end{center}
\caption[]{The percentage of charmonium production from color octets in 
NRQCD at each energy we consider.}
\label{octfrac}
\end{table}
The octet contribution decreases with energy for all charmonium states.
Since color singlet $\chi_{cJ}$ production is through the $gg$ and $gq$ 
channels, the fraction of octet $\chi_{cJ}$ production is quite small, 10\%
or less.  The decrease of octet production
with energy is expected because the octet $q \overline 
q$ contribution becomes even smaller at higher energies.  On the other hand,
direct $J/\psi$ and $\psi'$ production is color octet dominated since both
the $gg$ and $q \overline q$ channels have octet contributions, see 
Eq.~(\ref{psigg}).  The larger value of $\Delta_8(J/\psi)$ relative to
$\Delta_8(\psi')$ increases the 
overall octet contribution from $\sim 66$\% for the $\psi'$ to $\sim 87$\%
for the direct $J/\psi$ at $\sqrt{s} = 29.1$ GeV.  However, when the 
$\chi_{cJ}$ radiative decays are included, the octet contribution to 
total $J/\psi$ production is nearer to that of the $\psi'$, $\sim 63$\%.  
These results are
reflected in Fig.~\ref{nrqcd}.  Note also that the $x_F$ distributions of
the charmonium states are not exactly parallel to each other, as predicted
by the CEM.  
Unfortunately the slopes of the $x_F$ distributions are quite similar and it
is not until relatively large values of $x_F$ that the differences become more
significant.  However, there are other ways to distinguish the production
mechanism since these two models of charmonium production lead to
quite different predictions of the $A$ dependence, as we will demonstrate in
the next section.

\section{Absorption by Nucleons}

In Ref.~\cite{KSghfit}, absorption was described in terms of the singlet and
octet components of the $J/\psi$ wavefunction,
\be
|J/\psi \rangle = a_0 |(c \overline c)_1 \rangle + a_1 |(c \overline c)_8 g
\rangle + a_2 | (c \overline c)_1 gg \rangle + a_2' | (c \overline c)_8 gg
\rangle + \cdots \, \, .
\label{fockexp}
\ee
In the CSM \cite{baru1,baru2}, only the first component is nonzero
for direct $J/\psi$ production.  The $c \overline c$ pairs then pass 
through nuclear matter in small color singlet states and reach their
final state size outside the nucleus, at least when $x_F > 0$.
If $c \overline c$ pairs are predominantly produced in
color octet states, then it is the $|(c \overline c)_8 g \rangle$ state that
interacts with nucleons.  After the color octet $c \overline c$ is produced,
it can neutralize its color by a nonperturbative interaction with a gluon.
This octet state is fragile so that a gluon exchanged between it and a nucleon
would separate the $(c \overline c)_8$ from the gluon, exposing its color 
and, since the octet is unbound, break it up \cite{KSghfit}.  If the $|(c 
\overline c)_8 g \rangle$ state is free to evolve without
interaction, such as in $pp$ collisions, the additional gluon would be 
absorbed by the octet $c \overline c$ pair, hence `evaporating' the color.
The CEM does not then care about the relative coefficients in
Eq.~(\ref{fockexp}).  As formulated in Ref.~\cite{benrot}, the
NRQCD model provides the leading
coefficients in the expansion of the wavefunction
in Eq.~(\ref{fockexp}) and hence encompasses both singlet and octet production
and absorption.  In this section, we will describe the absorption of 
color singlets,
color octets, and the combination of the two for final-state $J/\psi$, $\psi'$
and $\chi_c$ production.  
Any differences in the $A$ dependence of these states will
be a consequence of this nucleon absorption.
We calculate charmonium production in the CEM with pure octet
and pure singlet absorption while NRQCD is used to determine the fraction of
charmonium states production in color singlets and color octets.  This then
determines the rate of singlet and octet absorption in Eq.~(\ref{fockexp}).

The effect of nuclear absorption alone on the $J/\psi$ 
production cross section in
$pA$ collisions may be expressed as \cite{rvrev} \be \sigma_{pA} =
\sigma_{pN} \int d^2b \,  \int_{-\infty}^{\infty}\, dz \, \rho_A (b,z) 
S^{\rm abs}(b,z)
\label{sigfull} \ee where $b$ is the impact parameter and $z$ is the
longitudinal production point.  When the production and absorption can be
factorized, as in the CEM, and no other $A$ dependent effects are included,
$\sigma_{pN}$ is independent of $A$ and drops out of the calculation of
$\alpha$.  The nuclear absorption survival probability,
$S^{\rm abs}$, is \be S^{\rm abs}(b,z) = \exp \left\{
-\int_z^{\infty} dz^{\prime} \rho_A (b,z^{\prime}) \sigma_{\rm abs}(z^\prime
-z)\right\} \label{nsurv} \ee 
The nucleon absorption cross section, $\sigma_{\rm abs}$, depends on where the
state is produced and how far it travels through nuclear matter.
Nuclear charge density
distributions from data are used for $\rho_A$ \cite{JVV}. The effective $A$
dependence is obtained from Eqs.~(\ref{sigfull}) and (\ref{nsurv}) by
integrating over $z'$, $z$, and $b$.  The full dependence on $A$ can be related
to $\alpha(x_F)$ in Eq.~(\ref{alfint}) but $\alpha$ is only constant if
$\sigma_{\rm abs}$ is constant and independent of the production mechanism
\cite{RVPRC,rvrev}.
The observed $J/\psi$ yield includes an $\approx 30$\% contribution from 
$\chi_{cJ}$ decays \cite{Teva} and an $\approx 12$\% contribution from   
$\psi'$ decays \cite{HPC}.  Then the total $J/\psi$ survival probability is
\be S_{J/\psi}^{\rm abs}(b,z) = 0.58 S_{J/\psi, \, {\rm dir}}^{\rm abs}(b,z) 
+ 0.3 S_{\chi_{cJ}}^{\rm abs}(b,z) + 0.12 S_{\psi'}^{\rm abs}(b,z) 
\, \, . \label{psisurv} \ee
The $\psi'$ and $\chi_c$ states are only produced directly since
other, more massive, charmonium resonances lie above the $D \overline D$
threshold and decay to $D \overline D$
pairs.  

We will present calculations for the total and direct $J/\psi$, 
$\psi'$, and $\chi_{cJ} \rightarrow J/\psi$ $A$ dependence.  
We include the $\chi_{cJ}$ branching
ratios to $J/\psi$ because even though the $\chi_{c0}$ cross section is large,
the small branching ratio gives it a negligible contribution 
to the final-state $J/\psi$ yield.
Our results will be calculated at 158, 450, and 920 GeV,
corresponding to the NA60 and HERA-B energies respectively.  We calculate
$\alpha(x_F)$ for two targets in each experiment: Be and Pb for NA60; C and W
for HERA-B.

\subsection{CEM: color singlet absorption}

We first discuss pure color singlet absorption.  In this case, $\sigma_{\rm
abs}$ depends on the size of the $c \overline c$ pair as it traverses the
nucleus.  This was first described in terms of color transparency \cite{bm}.
The \dd{c}{c} pairs are
initially produced with a size on the
order of its production time, $r_{\rm init} \sim \tau_{\rm init} 
\propto m_c^{-1}$.  This initial size is
ignored in the calculation.   
The charmonium formation time obtained from potential models \cite{KMS} is
$\tau_C \sim 1-2$ fm, 
considerably longer.  The absorption cross section of these small color singlet
pairs grows as a function of proper time until
$\tau_C$ 
when it saturates at its asymptotic value $\sigma_{C N}^{\rm s}$ 
\cite{RVPRC,GV,Blol3}, 
\be
\sigma_{\rm abs}(z^\prime -z) = \left\{ \begin{array}{ll} \sigma_{C 
N}^{\rm s} 
\bigg( {\displaystyle \frac{\tau}{\tau_C}} \bigg)^2 \,\, \, \, \,
\, \, & \mbox{if $\tau<\tau_C$} \\ \sigma_{C N}^{\rm s} 
& \mbox{otherwise}
\end{array}     \right. \, \, . \label{sigsing}
\ee 
The proper time $\tau$ is related to the path length
through nuclear matter by $\tau = (z^\prime - z)/
\gamma v$ where the $\gamma$ factor introduces $x_F$ and energy
dependencies to $\sigma_{\rm abs}$.  At low energies and negative
$x_F$, the \dd{c}{c} pair may hadronize
inside a large nucleus.  

\begin{figure}[htb]
\setlength{\epsfxsize=\textwidth}
\setlength{\epsfysize=0.55\textheight}
\centerline{\epsffile{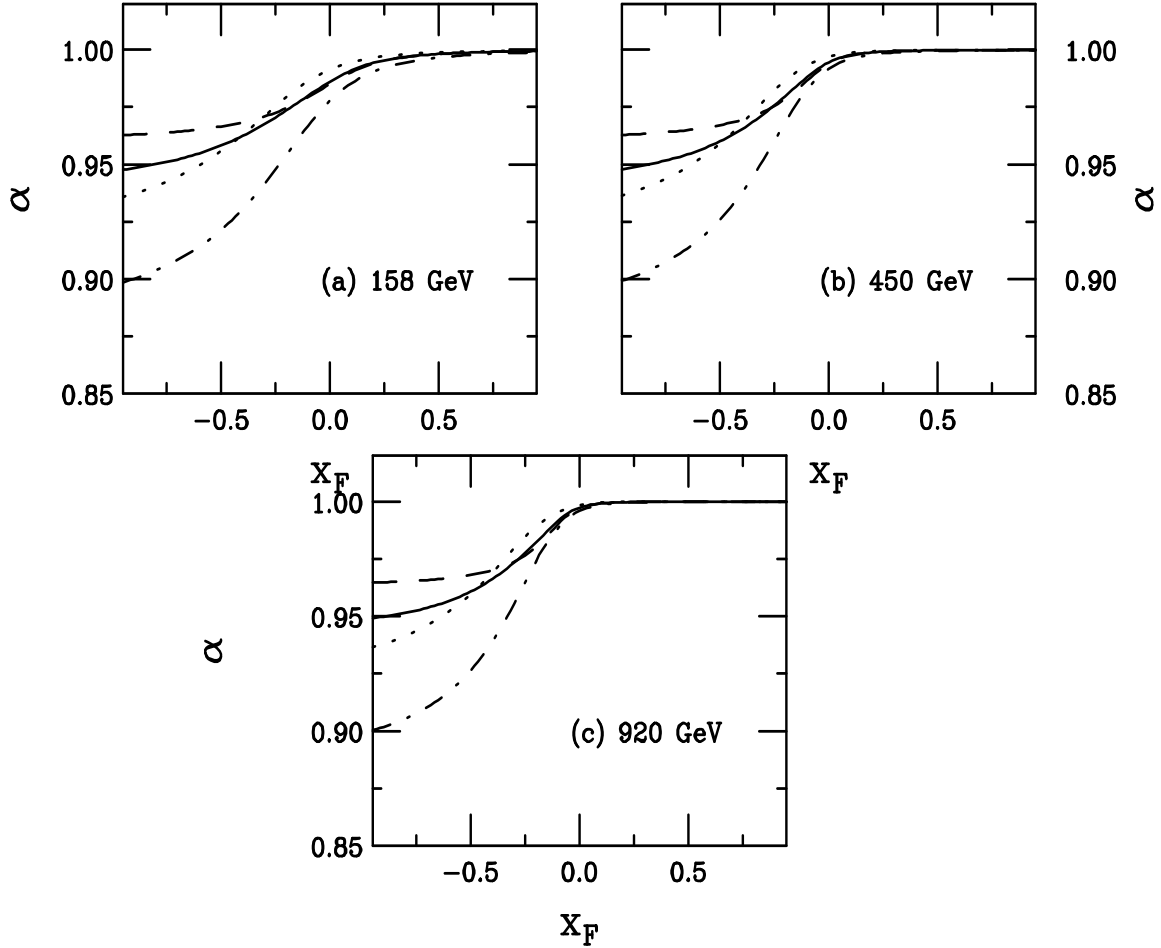}}
\caption[]{The $A$ dependence for color singlet absorption is shown.  The 
results are calculated at 158 GeV (a), 450 GeV (b), and 920 GeV (c).
The total $J/\psi$ (solid), the direct $J/\psi$ (dashed), the 
$\psi'$ (dot-dashed) and the $\chi_c$ (dotted)  $A$ dependencies are given.} 
\label{abssing}
\end{figure}

Figure~\ref{abssing} illustrates the energy dependence of color singlet
absorption in $pA$ interactions.  We take
$\sigma^{\rm s}_{J/\psi N} = 2.5$ mb \cite{KSoct}.
Assuming that the asymptotic absorption cross sections scale in
proportion to the squares of the
charmonium radii \cite{hpov},  we have $\sigma_{\psi' N}^{\rm s} \approx
3.7 \sigma_{J/\psi N}^{\rm s}$ and   $\sigma_{\chi_{cJ} N}^{\rm s} \approx
2.4 \sigma_{J/\psi N}^{\rm s}$.  Thus each contribution to Eq.~(\ref{psisurv})
has a different $A$ dependence.
The charmonium formation times are:
$\tau_{J/\psi} = 0.9$ fm, $\tau_{\psi'} = 1.5$ fm, and $\tau_{\chi_c} = 2.0$ 
fm 
\cite{KMS}.
The results at $x_F < 0$
reflect the differences in formation times as well as the change in the
$\gamma$ factor due 
to their masses.  The shorter formation time of the $J/\psi$ allows it to 
reach its asymptotic size at large negative $x_F$. 

At the lowest energy, 158 GeV, the charmonium states have a small chance of
being formed inside the target at $x_F > 0$ since $\alpha \neq 1$ at $x_F
\sim 0.2$ although the deviation from unity is small.
For higher energies, the charmonium
states are produced outside the nucleus for $x_F > 0$ so that 
$\alpha \approx 1$.  No observable differences appear between the charmonium
states at positive $x_F$.  Indeed, this $A$ dependence is in 
contradiction with all available data at $x_F \approx 0$ unless other nuclear 
effects are included \cite{RVPRC}.  Therefore this picture of absorption is
primarily useful for the interpretation of our calculations of pure octet
absorption in the CEM and the combination of singlet and octet production and
absorption in NRQCD.  

The direct $J/\psi$ $A$ dependence (dashed curve) 
is weakest because its asymptotic cross 
section is smallest.  The $\psi'$ $A$ dependence (dot-dashed curve)
is strongest because its 
final-state size and corresponding $\sigma_{\rm abs}$ is largest.  In the
calculation of Ref.~\cite{KMS}, the $\chi_c$ radius is somewhat smaller than
that of the $\psi'$ so that $\sigma^{\rm s}_{\psi' N} > \sigma^{\rm 
s}_{\chi_c N}$.  The
$\chi_c$ formation time is the longest of the charmonium states and thus most
likely to be produced outside the target.  Therefore the $\chi_c$ 
$\alpha$ is actually slightly larger than that of the direct $J/\psi$ at
$x_F \sim 0$ due to the longer $\chi_c$ formation time (dotted curve).  The 
$\chi_c$ contribution to the $A$ dependence of the
total $J/\psi$ yield (solid curve) decreases the total $J/\psi$
$\alpha$ at large negative $x_F$, more like the $\chi_c$, while when $x_F
\rightarrow 0$, the $\chi_c$ $\alpha$ is near unity and the total and direct
$J/\psi$ $A$ dependencies are the same.

\subsection{CEM: color octet absorption}

On the other hand, if the $c \overline c$ pairs are produced only in 
color octet states, they should hadronize
after $\tau_8 \sim 0.25$ fm in the $c \overline c$ rest frame \cite{KSoct}.  
At large $x_F$ in the lab frame, hadronization then occurs after
the $c \overline c$ has passed through the target as an octet.  These fast $c
\overline c$ pairs thus remain color octets until after they have left 
the nucleus.  However,
at negative $x_F$ it is possible for the octet states to neutralize their color
inside the nucleus and interact as color singlets during the remainder of their
path through the target \cite{KSoct}.  
This effect has typically been neglected when
studying the $A$ dependence of quarkonium production because the effect remains
small in the $x_F$ regions so far covered, $x_F > -0.1$ \cite{RVPRC,rvrev}.  
(See however Ref.~\cite{KSoct}.)  While traveling
through the nucleus as a pre-resonant $|(c \overline c)_8 g \rangle$ state,
the eventual identity of the final-state resonance is undetermined and all
quarkonium states are absorbed with the same cross section, $\sigma_{\rm 
abs}^{\rm o}$.  This physical picture agrees rather well with the empirical
evidence that the $J/\psi$ and $\psi'$ $A$ dependencies are similar 
over the measured 
$x_F$ range \cite{E772,e866}.  We choose $\sigma_{\rm abs}^{\rm o} = 3$ mb to
agree with $\alpha \approx 0.95$ for the $J/\psi$
measured by the E866 collaboration
at $x_F = 0$ \cite{e866} when no other nuclear effects are considered.
Note that this value is somewhat smaller than typically assumed for the color
octet cross section \cite{KSghfit} due to the relatively large measured
$\alpha$. 

We account for color
neutralization of the octet in the nucleus in the following way:  
The path length of the $| (c
\overline c)_8 g \rangle$ through the nucleus is calculated in the nuclear
rest frame.  If it exceeds the maximum path length through the nucleus from
the $| (c \overline c)_8 g \rangle$ production point, 
$\sigma_{\rm abs}^{\rm o} 
= 3$ mb for all charmonium states.  This is 
the case for $x_F \geq 0$ with all three energies.  
However, if color neutralization occurs before
the state escapes the target, the resulting color singlet is
absorbed according to Eq.~(\ref{sigsing}).  

\begin{figure}[htb]
\setlength{\epsfxsize=\textwidth}
\setlength{\epsfysize=0.55\textheight}
\centerline{\epsffile{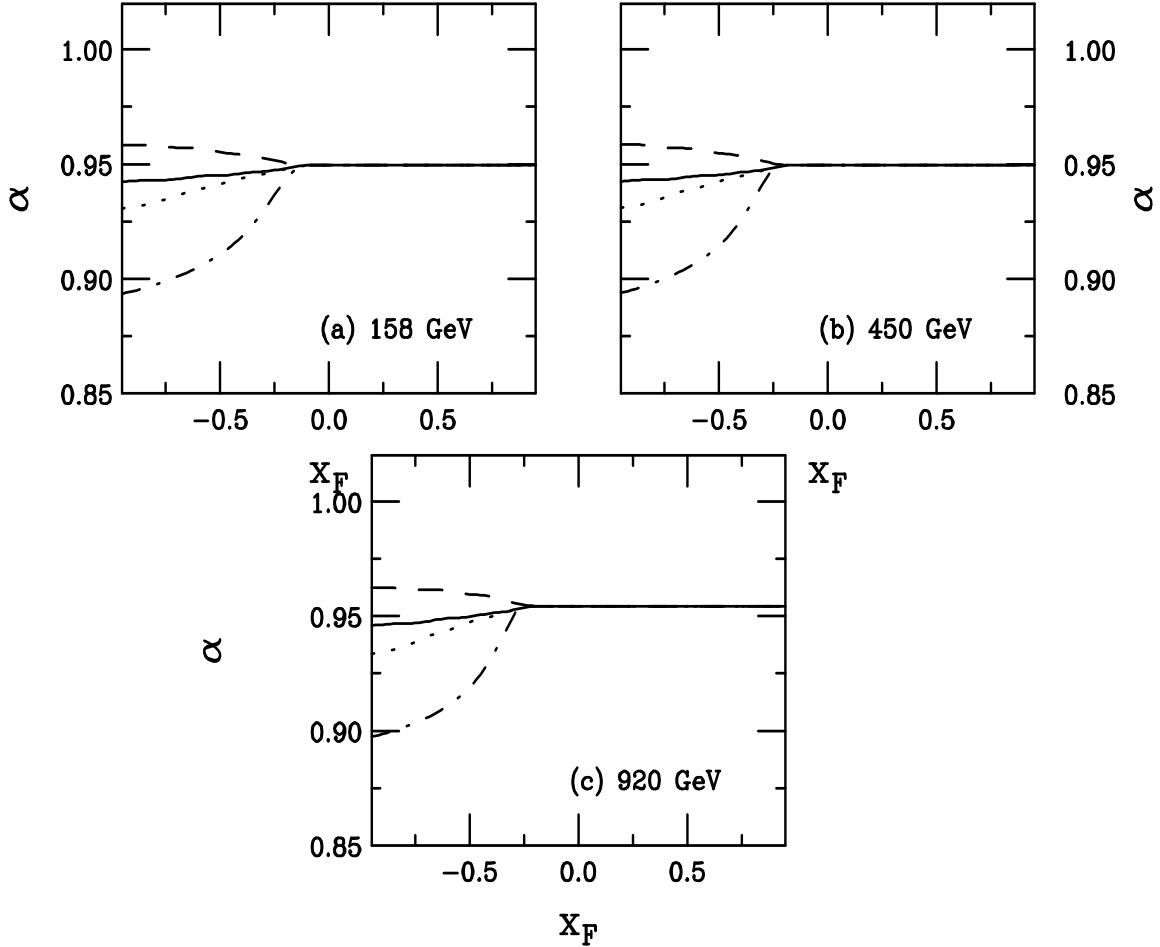}}
\caption[]{The $A$ dependence for color octet absorption is shown.  The 
results are calculated at 158 GeV (a), 450 GeV (b), and 920 GeV (c).
The total $J/\psi$ (solid), the direct $J/\psi$ (dashed), the 
$\psi'$ (dot-dashed) and the $\chi_c$ (dotted)  $A$ dependencies are given.} 
\label{absoct}
\end{figure}

The $A$ dependence of color octet absorption is shown 
in Fig.~\ref{absoct}.  Note
that at 158 GeV, color neutralization is achieved for $x_F \leq -0.2$.  At
higher energies, neutralization occurs in the target
at larger negative $x_F$.  It is important to remember that just 
because the octet color has been neutralized, the
asymptotic cross section is not necessarily 
reached inside the target.  With a formation time
of less than 1 fm, only the $J/\psi$ is likely to be fully formed in a large
nucleus, as observed in the `saturation' of the $A$ dependence at $x_F \leq
-0.5$.  On the other hand, although the $\psi'$ and $\chi_c$ may
become singlets
inside the target, they do not reach their final-state size inside the target,
even at $x_F \rightarrow -1$, due to the combination of their larger radii and
longer formation times.  The $\psi'$ and $\chi_c$ $A$ dependencies thus do not
saturate, even at low energies.  The slightly higher $\alpha$ at 920 GeV
is due to the different target $A$ ratios chosen for the NA60 and HERA-B 
calculations, Pb/Be and W/C respectively.

We point out that the calculated $\alpha$
is lower in the pure color octet 
picture at $x_F \rightarrow -1$ than in the color singlet
absorption model even though the same 
asymptotic color singlet cross sections are used.  This is 
because now the state starts out as a color octet with a finite probability to
be absorbed before neutralizing its color.  The probability tends to be larger
in real nuclei where the path length is calculated in the integral 
over impact parameter rather than in the 
empirical analytic model where an average path
length is used \cite{KSoct}.  The color octet is absorbed with its full cross
section which is larger than the color singlet cross section at the point of
absorption even though the asymptotic color singlet cross sections may be 
greater, {\it e.g.} $\sigma^{\rm o}_{\psi' N} < \sigma^{\rm s}_{\psi' N}$.
The effective octet absorption
cross section is larger because in the color octet state absorption has 
essentially no time delay.

The greatest differences in the $A$ dependencies of the states 
are at intermediate to
large negative $x_F$ and would be most easily observable by NA60 at 158 GeV
if their coverage extended so far.  Note that, in this case also, the total 
$J/\psi$ and the $\chi_c$ $A$ dependencies would be quite similar in the target
region while the $\psi'$ $\alpha$ would be lower, $\sim 0.9$ at 158 GeV and
$x_F = -0.5$ compared to $\sim 0.93$ for the
total $J/\psi$ and $\chi_c$.  However,
only HERA-B has the capability to measure the $A$ dependence at negative $x_F$
and at the higher energy the differences appear at higher negative 
$x_F$ and are generally not as large.  Thus any distinction 
will be rather difficult to determine
and the observed $A$ dependence is likely to be
the same within the experimental uncertainties for all charmonium states.

\subsection{NRQCD: color singlet and color octet absorption}

Recall that in the preceding discussion, only one type of $c \overline c$ color
state is assumed to be produced, either singlet or octet.  
Therefore the absorption factorizes from the
production mechanism and the CEM cross section cancels in the calculation of 
$\alpha$, as in Eq.~(\ref{sigfull}).  
The result is then independent of all parameters in the production
process such as $m_c$ and the parton densities.
However, according to Eq.~(\ref{fockexp}),
charmonium production is through a combination of octet and singlet
states.  In this case, production and absorption are intimately related
and the NRQCD cross section determines the relative octet proportion for each
state as a function of $x_F$ \cite{Zhang1,Zhang2}.  
The ratio of octet
to singlet production is energy and $x_F$ dependent, as shown in 
Table~\ref{octfrac} and Fig.~\ref{nrqcd}.  Therefore Eq.~(\ref{sigfull}) does
not hold since $\sigma_{pN}$ and $S^{\rm abs}$ do not factorize for all $x_F$.

We now give the unfactorized $x_F$ distributions for each state in NRQCD.
The $x_F$ dependence of direct charmonium production and absorption 
is straightforward:
\be 
\frac{d\sigma_{pA}^{\psi}}{dx_F} & = & \int d^2b \left[
\frac{d\sigma_{pp}^{\psi, \, \rm 
oct}}{dx_F} \, T_A^{\psi, \rm eff\, (oct)}(b) +
\frac{d\sigma_{pp}^{\psi, \, \rm sing}}{dx_F} \, T_A^{\psi,
\rm eff\, (sing)}(b) \right] \, \, , \label{psipcombo} \\
\frac{d\sigma_{pA}^{\chi_{cJ} \rightarrow J/\psi X}}{dx_F} & = & 
\int d^2b \sum_{J=0}^2 B(\chi_{cJ}
\rightarrow J/\psi X) \left[ \frac{d\sigma_{pp}^{\chi_{cJ}, \, \rm 
oct}}{dx_F} \, T_A^{\chi_{cJ}, \rm eff\, (oct)}(b) \right. \nonumber
\\ &  & \left. \mbox{} +
\frac{d\sigma_{pp}^{\chi_{cJ}, \, \rm sing}}{dx_F} \, 
T_A^{\chi_{cJ},\rm eff\, (sing)}(b) \right] \, \, , \label{chiccombo} 
\ee
where $\psi = J/\psi, \, \psi'$ and
$T_A^{\rm eff} = \int dz \rho_A S^{\rm abs}$ for both singlet and octet
absorption.  The $pp$ subscript is used to denote unmodified parton
distributions in the target.
The total $J/\psi$ $x_F$
distribution is more complex since it
includes the feeddown from the $\psi'$ and $\chi_c$ states.
Then \cite{RVPRC}
\be \lefteqn{\frac{d \sigma_{pA}^{J/\psi, \,{\rm tot}}}{dx_F} 
= \int d^2b \left\{ \left[
\frac{d \sigma_{pp}^{J/\psi, \, {\rm dir, \, oct}}}{dx_F} T_A^{J/\psi,
\rm eff\, (oct)}(b) \right. \right. } &  & \nonumber \\
& & \!\!\! \!\!\! \!\!\! 
\left. \mbox{} + \sum_{J=0}^2 B(\chi_{cJ} \rightarrow J/\psi X) \frac{d 
\sigma_{pp}^{\chi_{cJ}, \, {\rm oct}}}{dx_F} T_A^{\chi_{cJ},
\rm eff\, (oct)}(b) 
+ B(\psi' \rightarrow \psi X) \frac{d\sigma_{pp}^{\psi', \, 
{\rm oct}}}{dx_F} T_A^{\chi_{cJ}, \rm eff\, (oct)}(b)\right] 
\nonumber \\ & & \!\!\! \!\!\! \!\!\! \mbox{} + 
\left[ \frac{d \sigma_{pp}^{J/\psi, \, {\rm dir, \, sing}}}{dx_F} T_A^{J/\psi,
 {\rm dir}, {\rm eff\, (sing)}}(b)
+ \sum_{J=0}^2 B(\chi_{cJ} \rightarrow \psi X) \frac{d 
\sigma_{pp}^{\chi_{cJ}, \, {\rm sing}}}{dx_F} T_A^{\chi_{cJ},\,
 {\rm eff\, (sing)}}(b) \right. \nonumber \\ & & \!\!\! \!\!\! \!\!\! 
\left. \left. \mbox{}
+ B(\psi' \rightarrow \psi X) \frac{d\sigma_{pp}^{\psi', \, {\rm sing}}}{dx_F}
 T_A^{\psi', \, {\rm eff\, (sing)}}(b)
\right] \right\} \label{psicombo}
\, \, . \ee

\begin{figure}[htb]
\setlength{\epsfxsize=\textwidth}
\setlength{\epsfysize=0.55\textheight}
\centerline{\epsffile{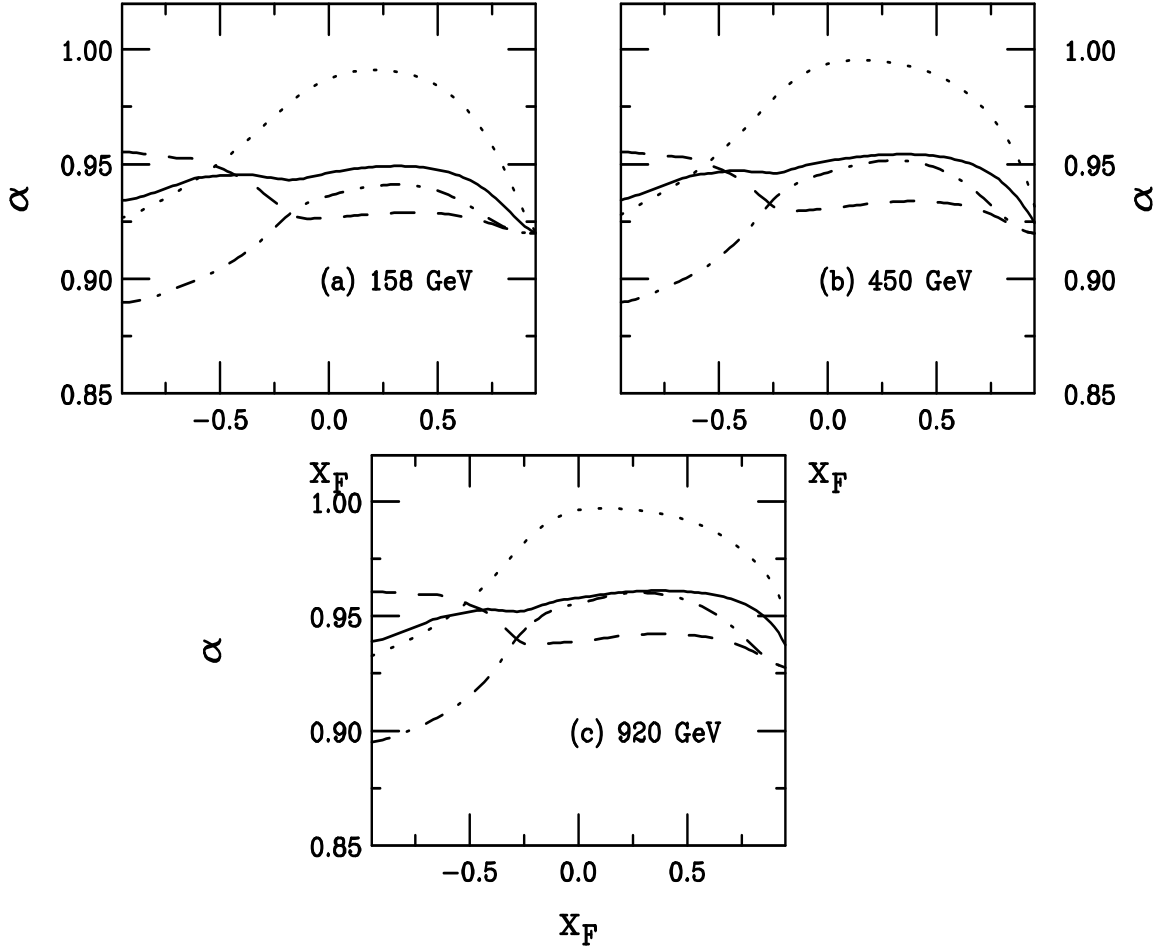}}
\caption[]{The $A$ dependence for color singlet and color octet absorption in
the NRQCD model is shown.  The 
results are calculated at 158 GeV (a), 450 GeV (b), and 920 GeV (c).
The total $J/\psi$ (solid), the direct $J/\psi$ (dashed), the 
$\psi'$ (dot-dashed) and the $\chi_c$ (dotted) $A$ dependencies are given.} 
\label{abscombo}
\end{figure}

Our results are shown in Fig.~\ref{abscombo}.   We have chosen the octet 
absorption cross section such 
that the total $J/\psi$ $\alpha$ agrees in magnitude 
with the recent measurement by E866 at 800 GeV \cite{e866}. 
In this case, $\sigma_{J/\psi N}^{\rm oct} 
= 5$ mb and $\sigma_{J/\psi N}^{\rm sing} = 2.5$ mb
gives $\alpha \approx 0.95$ at $x_F \approx 0$.
The same octet cross section 
is then used for the octet component of the absorption for all
charmonium states.  If the octet state neutralizes its color, the resulting 
color singlet is absorbed according to Eq.~(\ref{sigsing}).  The 
same absorption cross sections are also used for the singlet 
component of charmonium production in section 3.1.  

In Figs.~\ref{abssing} and \ref{absoct},
the predicted difference between $J/\psi$, $\psi'$ and $\chi_c$ absorption
was not large in the measurable region, particularly when the total $J/\psi$
$A$ dependence
was considered.  Now, however, the $\chi_c$ and total $J/\psi$ results are
significantly different and if the NRQCD model provides the right description 
of charmonium production, the measured $\chi_c$ $A$ dependence should thus
be quite
different from that of the $J/\psi$.  Note also that the $\psi'$ $\alpha$ is 
slightly lower than the total $J/\psi$ $\alpha$ 
at $x_F \sim 0$, in accordance with the
E866 results \cite{e866}.

The direct $J/\psi$ $A$ dependence is rather similar to the octet results
shown in Fig.~\ref{absoct} due to its large octet component.  The $\psi'$
has a larger overall singlet component but the singlet 
influence on the $A$ dependence
is rather weak.  The main difference between direct $J/\psi$ and $\psi'$
production at $x_F > 0$  is the larger $\alpha$ of the $\psi'$ due to the 
singlet component.  However, the dominant color singlet component of $\chi_c$
production leads to an almost linear $A$ dependence for $0< x_F < 0.5$ at
158 GeV and $-0.25 < x_F < 0.5$ at 920 GeV.  The range of $x_F$ at which
$\alpha \sim 1$ is broader at higher energies because the singlet $gg$
contribution grows larger with energy.  Given the similarities between the
pure color singlet and color octet results at large negative $x_F$, it is
difficult to disentangle the relative contributions for the 
combination of the two in this region.  
However, the differences at large positive $x_F$ are
due to the change in the relative octet/singlet contributions.  At large $x_F$,
the $q \overline q$ component is more important.  This octet piece causes the
drop in $\alpha$ of the $\chi_c$ at large $x_F$ while having little effect on
the $J/\psi$ and $\psi'$.  Finally, we note that the 
total $J/\psi$ $A$ dependence in this calculation is quite similar to the
$\psi'$ dependence, as already indicated by previous measurements 
\cite{E772,e866}.

It is clear that if this model is correct, both NA60 and HERA-B should have
no difficulty observing substantial differences in the $J/\psi$ and $\chi_c$
$A$ dependence since the values of $\alpha$ are clearly different even at
positive $x_F$.  Other effects such as nuclear shadowing and energy loss would
be similar for the two resonances so that differences in 
absorption mechanisms would not be washed out.  

\section{Conclusions}

We have calculated the nuclear dependence of total and direct $J/\psi$, $\psi'$
and $\chi_c$ due to absorption alone.  We have studied absorption of pure color
singlets and color octets in the context of the color evaporation model and
a combination of octet and singlet production in nonrelativistic QCD.  When
considering charmonium production in a pure color state, as in the color
evaporation model, we find little
difference in the charmonium $A$ dependencies in regions accessible to past 
experiments, in agreement with the $J/\psi$ and $\psi'$ measurements to date.
However, when the $\chi_c$ is considered, its large color singlet component
results in a substantially different $A$ dependence in the
nonrelativistic QCD description.  This difference should be 
easily detected by the two
experiments that plan to measure $\chi_c$ production, NA60 and HERA-B.
Their results should quickly answer the question posed by the title of this 
paper.

{\bf Acknowledgments}
I would like to thank M. Bruinsma, D. Hansen, C. Louren\c{c}o, M. Medinnis, 
K. Redlich, H. Satz and A. Zoccoli for
helpful discussions.

\end{document}